\documentclass[english, twocolumn, 10pt, aps, superscriptaddress, floatfix, prb, citeautoscript]{revtex4-1}
\pdfoutput=1
\usepackage[utf8]{inputenc}
\usepackage[T1]{fontenc}
\usepackage{verbatim}
\usepackage{units}
\usepackage{mathtools}
\usepackage{amsmath}
\usepackage{amssymb}
\usepackage{graphicx}
\usepackage{wasysym}
\usepackage{layouts}
\usepackage{siunitx}
\usepackage{bm}
\usepackage{xcolor}
\usepackage[colorlinks, citecolor={blue!50!black}, urlcolor={blue!50!black}, linkcolor={red!50!black}]{hyperref}
\usepackage{bookmark}
\usepackage{tabularx}
\usepackage{microtype}
\usepackage{babel}
\hypersetup{pdfauthor={Quantum Tinkerer},pdftitle={Enhanced proximity effect in zigzag-shaped Majorana Josephson junctions.}}

\newcommand{\kx}{k_x}
\newcommand{\ky}{k_y}
\newcommand{\meff}{m_\text{eff}}

\renewcommand{\comment}[2]{#2}
\DeclarePairedDelimiter\abs{\lvert}{\rvert}
\DeclarePairedDelimiter\norm{\lVert}{\rVert}

\makeatletter
\let\oldabs\abs
\def\abs{\@ifstar{\oldabs}{\oldabs*}}
\let\oldnorm\norm
\def\norm{\@ifstar{\oldnorm}{\oldnorm*}}
\makeatother

\newcolumntype{L}[1]{>{\raggedright\arraybackslash}p{#1}}
\newcolumntype{C}[1]{>{\centering\arraybackslash}p{#1}}
\newcolumntype{R}[1]{>{\raggedleft\arraybackslash}p{#1}}

\begin{document}

\title{Enhanced proximity effect in zigzag-shaped Majorana Josephson junctions.}

\author{Tom Laeven}
\email[Electronic address: ]{tlaeven@hotmail.com}
\affiliation{Kavli Institute of Nanoscience, Delft University of Technology, P.O. Box 4056, 2600 GA Delft, The Netherlands}
\author{Bas Nijholt}
\email[Electronic address: ]{bas@nijho.lt}
\affiliation{Kavli Institute of Nanoscience, Delft University of Technology, P.O. Box 4056, 2600 GA Delft, The Netherlands}
\author{Michael Wimmer}
\email[Electronic address: ]{m.t.wimmer@tudelft.nl}
\affiliation{Kavli Institute of Nanoscience, Delft University of Technology,
 P.O. Box 4056, 2600 GA Delft, The Netherlands}
\affiliation{QuTech, Delft University of Technology,
 P.O. Box 4056, 2600 GA Delft, The Netherlands}
\author{Anton R. Akhmerov}
\email[Electronic address: ]{zigzag@antonakhmerov.org}
\affiliation{Kavli Institute of Nanoscience, Delft University of Technology, P.O. Box 4056, 2600 GA Delft, The Netherlands}

\date{2019-03-14}
\begin{abstract}
High density superconductor-semiconductor-superconductor junctions have a small induced superconducting gap due to the quasiparticle trajectories with a large momentum parallel to the junction having a very long flight time.
Because a large induced gap protects Majorana modes, these long trajectories constrain Majorana devices to a low electron density.
We show that a zigzag-shaped geometry eliminates these trajectories, allowing the robust creation of Majorana states with both the induced gap $E_\textrm{gap}$ and the Majorana size $\xi_\textrm{M}$ improved by more than an order of magnitude for realistic parameters.
In addition to the improved robustness of Majoranas, this new zigzag geometry is insensitive to the geometric details and the device tuning.
\end{abstract}

\maketitle

%%██████████████████████████████████████████████████████████████████████████
%%██ Introduction
%%██████████████████████████████████████████████████████████████████████████
\section{Introduction}
\comment{Hybrid NS structures become topological and are useful for TQC.}
A hybrid structure containing a semiconductor with strong spin-orbit coupling coupled to a superconductor can become topological upon application of a magnetic field stronger than a critical field $B_\textrm{crit}$, with Majorana bound states appearing on its edges~\cite{Lutchyn2010,Oreg2010}.
Majorana bound states are a promising candidate to form the basis of a stable platform for topological quantum computing~\cite{Alicea2012,Beenakker2013,Beenakker2016,Leijnse2012}.
Much of the experimental effort~\cite{Mourik2012,Das2012,Deng2012,Churchill2013,Zhang2018} currently focuses on creating pairs of Majorana bound states in hybrid normal-superconductor (NS) nanowire structures.

\comment{SNS junctions also work and require smaller field.}
Recently, a modified setup has been proposed\cite{Pientka2017,Hell2017} relying on a superconductor-normal-superconductor (SNS) junction to lower the critical magnetic field $B_\textrm{c}$ by introducing a superconducting phase difference $\phi$.
When both NS interfaces are transparent the SNS junction enters the topological phase at $\phi=\pi$ at any finite $B$ field.
Two groups~\cite{Fornieri2018,Ren2018} have realized this system experimentally, but did not yet observe a hard induced superconducting gap.

\begin{figure}[!htb]
\centering
\includegraphics[width=0.9\columnwidth]{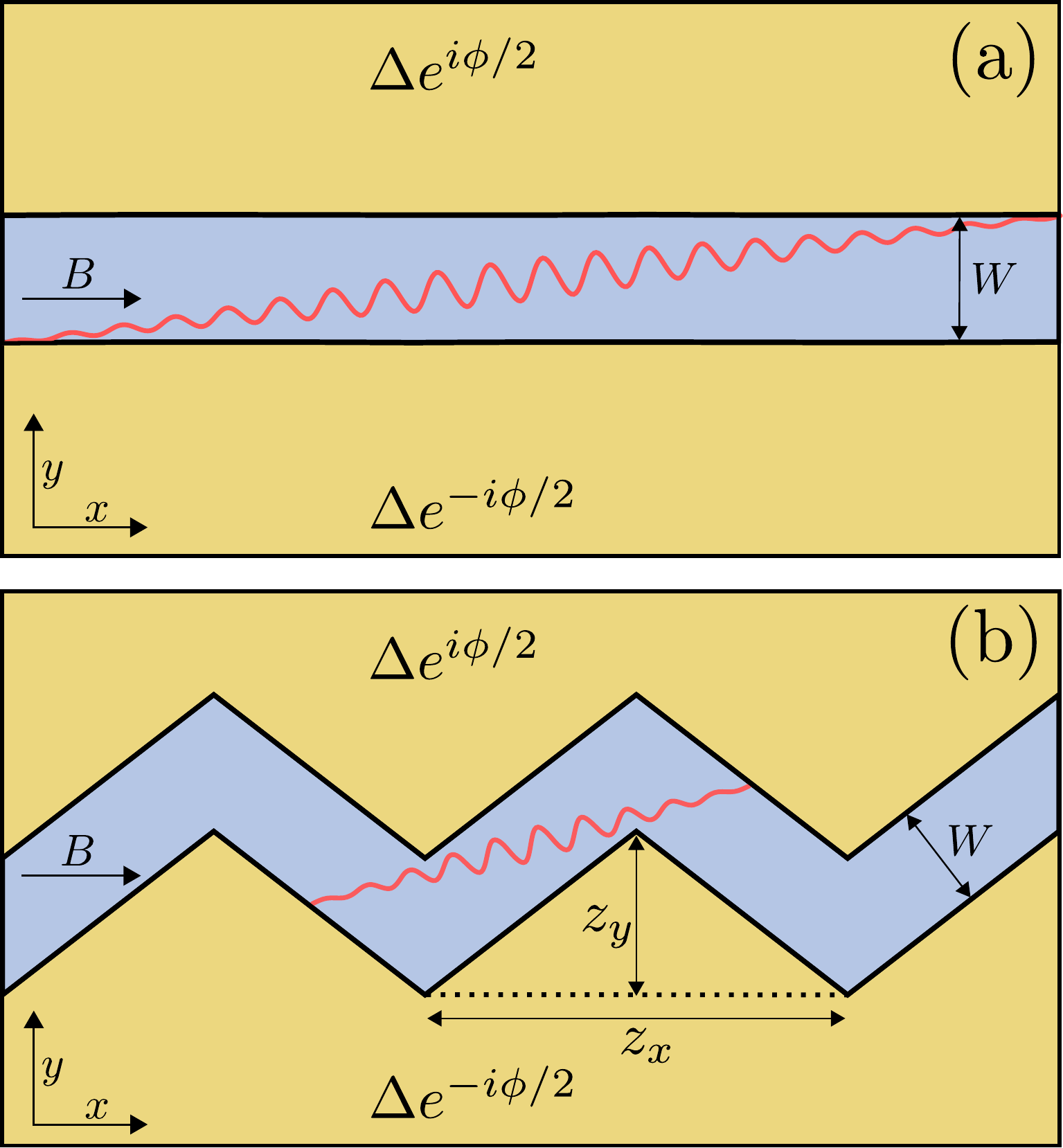}
\caption{The straight (top) and the zigzag (bottom) SNS junction.
The zigzag pattern has a peak-to-peak amplitude $z_y$ and a period $z_x$.
The yellow areas are superconductors with a phase difference of $\phi$ between the top and the bottom.
The middle area is the semiconductor of width $W$.
A magnetic field $B$ pointing in the $x$-direction causes a Zeeman splitting in the semiconductor
A trajectory traveling at a grazing angle (red curve) has a very long flight time $\tau_\textrm{f}$ and a very small induced gap $E_\textrm{gap} \ll \Delta$.
At the same time, the zigzag geometry limits the length of a trajectory therefore lowering $\tau_\textrm{f}$ and increasing $E_\textrm{gap}$.
\label{fig:setup}}
\end{figure}

\comment{The gap is small because of long trajectories.}
An important challenge in creating stable Majoranas is the appearance of a soft gap---a power law decay instead of an exponential decay of the density of states near zero energy.
In clean systems soft gap arises due to the reduction of the induced gap for states with the momentum directed along the junction~\cite{Gennes1963,Beenakker2005}.
From a semiclassical perspective, these momenta correspond to long paths through the semiconductor without interruption by the superconductor, shown in Fig.~\ref{fig:setup}(a).
These long trajectories have long flight times $\tau_\textrm{f} \approx L_\textrm{t} / v_F$ (see Fig.~\ref{fig:setup}), where $L_\textrm{t}$ is the trajectory length.
Equivalently, the Thouless energy of these trajectories $E_{\textrm{Th}}=\hbar / \tau_\textrm{f}$ is small, resulting in a small gap $E_\textrm{gap} \ll \Delta$.
This problem does not appear when the Fermi surface is small and the zero point motion dominates the transverse velocity, making a low filling of the bands a possible workaround\cite{Beenakker2005,Nijholt2015}.
However, low filling requires precise knowledge of the system and is more sensitive to disorder or microscopic inhomogeneities.
On the other hand, disorder scatters these long trajectories and introduces a cutoff on the scale of the mean free path~\cite{Golubov1988,Belzig1996,Pilgram2000} which Ref.~\onlinecite{Haim2018} proposes to use to improve Majorana properties; however, disorder is impossible to control to a required precision experimentally.

\comment{We show that zigzag geometry solves this problem by eliminating the long trajectories.}
We propose a new experimental setup (see Fig.~\ref{fig:setup}(b)) for the creation of Majoranas that eliminates long trajectories and therefore prevents the appearance of a soft gap, while also increasing the topological gap (the smallest gap in the dispersion relation) by more than an order of magnitude, depending on the parameters.
The setup consists of a zigzag or snake-like geometry for the semiconductor where long trajectories are not possible due to the geometry.
In this paper we will focus on two-dimensional (2D) Josephson junctions; however, a zigzag geometry will also work with only one superconductor.\footnote{We confirmed this numerically, although it is not included in this paper.}

%%██████████████████████████████████████████████████████████████████████████
%%██ Setup
%%██████████████████████████████████████████████████████████████████████████
\section{Setup}\label{sec:setup}

\comment{We consider a two-dimensional system with zigzag and BdG Hamiltonian.}
We consider a Josephson junction (Fig.~\ref{fig:setup}) consisting of a 2D strip of semiconductor, with superconductors on both sides.
We modulate the shape of the normal region, which can be either zigzag as depicted [Fig.~\ref{fig:setup}(b)], or a more smooth sinusoidal-like shape.
Similar to the conventional straight system \cite{Pientka2017}, a magnetic field $B_x$ perpendicular to the junction is applied.
We model the system with a Bogoliubov-de Gennes Hamiltonian (BdG):
\begin{subequations}
\begin{align}
    H_\textrm{N} = & \left[\frac{\hbar^2\left(\kx^2 + \ky^2\right)}{2\meff} - \mu + \alpha \left( \ky \sigma_x - \kx \sigma_y \right) \right] \tau_z
        + E_\text{Z} \sigma_x, \\
    H_\textrm{SC} = & \left[\frac{\hbar^2\left(\kx^2 + \ky^2\right)}{2\meff} - \mu\right] \tau_z
        + \Delta \cos{\frac{\phi}{2}} \tau_x + \Delta \sin{\frac{\phi}{2}} \tau_y.
\end{align}
\label{eq:hamiltonian}
\end{subequations}
Here $H_\textrm{N}$ and $H_\textrm{SC}$ are the Hamiltonians of the semiconductor and superconductors, respectively.
The normal part has a linear Rashba spin-orbit coupling term with strength $\alpha$ and a Zeeman field with $E_\text{Z}=\frac{1}{2} \mu_B g B_x$.
The superconductor has a coupling term $\Delta$, and the phases of the superconductors equal to $\pm\phi/2$.
Both the normal part and the superconductors have a kinetic term and chemical potential $\mu$.
The BdG Hamiltonian acts on the spinor wave function $\Psi={\left(\psi_{e\uparrow},\psi_{e\downarrow},\psi_{\textrm{h}\downarrow},-\psi_{\textrm{h}\uparrow}\right)}^{T}$, where $\psi_e$, $\psi_\textrm{h}$ are its electron and hole components, and $\psi_\uparrow$, $\psi_\downarrow$ are the spin-up and spin-down components.
The Pauli matrices $\sigma_{i}$ act on the spin degree of freedom and $\tau_{i}$ act on the electron-hole degree of freedom.
We consider a zigzag pattern with a period $z_x$, a peak-to-peak amplitude $z_y$, and $W$ the width of the junction [see Fig.~\ref{fig:setup}(b)].
Later we relax this assumption and show that the exact shape is unimportant.

\comment{We discretize the Hamiltonian and simulate it with Kwant.}
We discretize our continuum Hamiltonian [Eq.~\eqref{eq:hamiltonian}] on a square grid and implement a tight-binding model using Kwant~\cite{Groth2014}.
To preferentially sample important regions of parameter space, we use the Adaptive package~\cite{Nijholt2019a}.
The entire source code and the resulting raw data are available in Ref.~\onlinecite{Nijholt2019}.

\comment{The default parameters are ...}
Unless noted differently, the Hamiltonian parameters are $\alpha=\SI{20}{\meV \nm}$, $g=26$, $\meff=\SI{0.02}{\electronmass}$, $\mu=\SI{10}{\meV}$, $B_x=\SI{1}{T}$, $\phi=\pi$, and $\Delta=\SI{1}{\meV}$; and the geometry parameters are $W=\SI{200}{\nm}$, the period of the zigzag $z_x=\SI{1300}{\nm}$, the discretization contant $a=\SI{10}{\nm}$, and the lengths of the superconductors $L_\textrm{SC}=\SI{300}{\nm}$.

%%██████████████████████████████████████████████████████████████████████████
%%██ Band stuctures
%%██████████████████████████████████████████████████████████████████████████
\section{Band stuctures}\label{sec:band_structures}

\begin{figure}[!htb]
\includegraphics[width=\columnwidth]{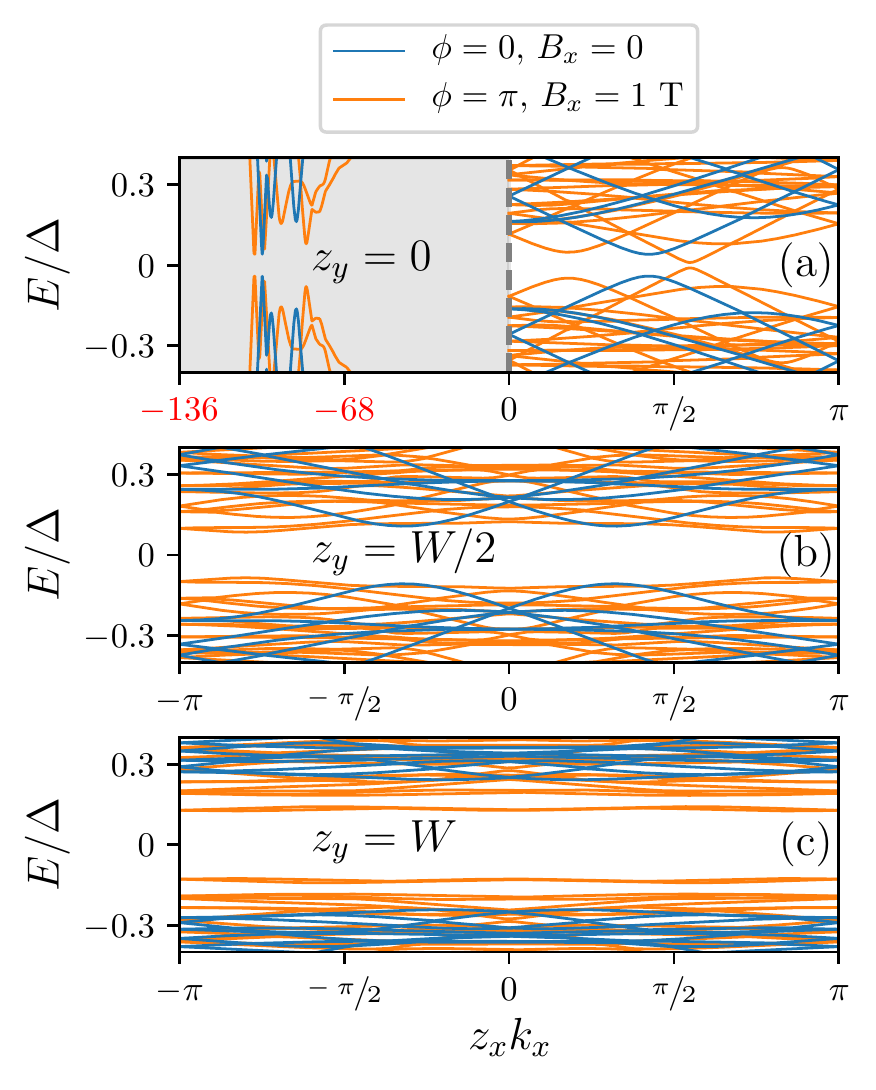}
\caption{Band stuctures of the system in Fig.~\ref{fig:setup}(b) with different zigzag amplitudes.
The blue lines correspond to a trivial phase ($\phi=0$, $B_x = 0$) and the orange lines to a topological phase ($\phi=\pi$, $B_x = \SI{1}{T}$).
The three subplots are for different amplitudes of the zigzag, with (a) a straight system $z_y=0$, (b) $z_y=W/2$, and (c) $z_y=W$, where $W=\SI{200}{\nm}$ is the junction width.
Subplot (a) has a different $x$-scale for $k_x < 0$ from the other subplots and displays the unfolded band structure.
For the right-hand side of (a) ($k_x > 0$), (b), and (c), the folding is the same, such that the velocity $v=dE/dk$ can be compared visually.
We observe that once there are no more straight trajectories inside the junction (when $z_y=W$) the spectrum becomes insensitive to the momentum $k_x$ and equivalently, $v_\textrm{F}$ decreases.
As the zigzag amplitude increases, the band gap $E_\textrm{gap}$ increases by an order of magnitude.
The combination of these ensures a significant decrease of the Majorana size because $\xi_M \propto v_\textrm{F}/E_\textrm{gap}$
The parameter values are written at the end of Sec.~\ref{sec:setup}.\label{fig:band_structures}}
\end{figure}

\comment{We calculate the band structure for varying amount of zigzag.}
We apply sparse diagonalization to the supercell Hamiltonian at different momenta $k_x$ to compute the band structure.
Because of the large periodicity of the zigzag and the resulting large supercell, the band structure is heavily folded.
In Fig.~\ref{fig:band_structures} we show the resulting band structures of zigzag systems with varying $z_y$.
The introduction of the zigzag has a striking effect: the bands flatten out and the topological gap increases by more than an order of magnitude.

\comment{Zigzag improves the gap and size because of cutting of trajectories and increasing transparency.}
In the unfolded band structure of a straight system, shown in Fig.~\ref{fig:band_structures}(a), the lowest energy states occur at $k \approx k_F$.
We interpret the increase of the gap $E_\textrm{gap}$ shown in Fig.~\ref{fig:band_structures}(b) and (c) as an effect of the zigzag geometry removing these long trajectories traveling at grazing angles.
Besides the increased $E_\textrm{gap}$, the states from different segments of the zigzag pattern have a negligible overlap and therefore have a vanishing velocity.
This reduction in velocity strongly reduces the Majorana size, as we discuss in section \ref{sec:shape_effects}.
Finally, in a zigzag geometry, every trajectory encounters a superconductor close to normal incidence.
Normal incidence has a higher transmission probability for entering the superconductor and therefore a higher Andreev reflection amplitude.
This provides another mechanism of the gap enhancement.

%%██████████████████████████████████████████████████████████████████████████
%%██ Localization lengths and shape effect
%%██████████████████████████████████████████████████████████████████████████
\section{Localization lengths and shape effects}\label{sec:shape_effects}

\begin{figure}[!htb]
\includegraphics[width=\columnwidth]{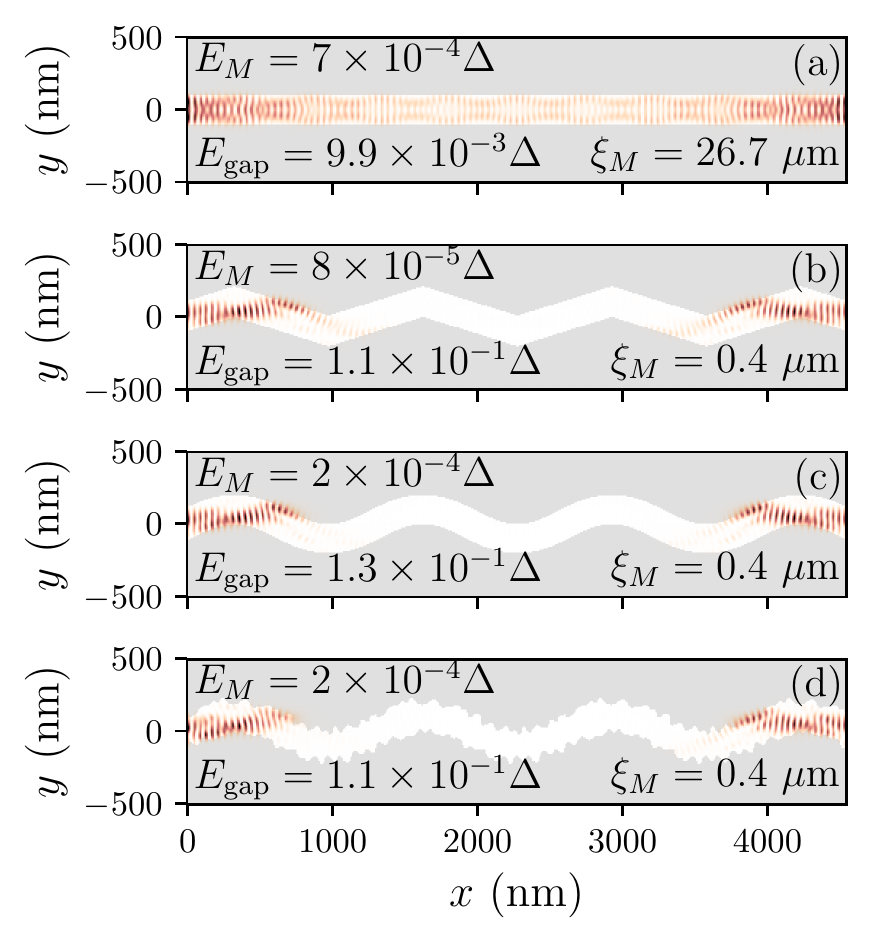}
\caption{Density of Majorana wave functions $\left| \psi_\textrm{M} \right|^2$ for sizes and geometries.
With (a) a straight system, (b) a zigzag system, (c) a system where lines parallel to a sinusoid defines the normal region, and (d) similar to (c) but with disordered edges.
Inside the figure, we indicate the Majorana length (or coherence length) $\xi_\textrm{M}$, the Majorana energy $E_\textrm{M}$ (the energy of the first excited state), and the topological energy gap $E_\textrm{gap}$.
We observe that $\xi_\textrm{M}$ for the straight system is almost two orders of magnitude longer and $E_\textrm{gap}$ more than an order of magnitude smaller than for the zigzag systems.
The robustness of $E_\textrm{gap}$ and $\xi_\textrm{M}$ across the zigzag geometries means that the details of the geometry do not matter for the improvements to occur.
The length of the system is $3.5 z_x=\SI{4550}{\nm}$, the remaining parameter values are written at the end of Sec.~\ref{sec:setup}.\label{fig:wave_functions}}
\end{figure}

\comment{We calculate the wave functions and find the Majorana lengths by fitting an exponential.}
We model a finite system and compute the Majorana wave function in different geometries: ribbon, zigzag, sine-like parallel curves, and a variant of the latter with disordered edges.
By diagonalizing the Hamiltonian, we find the Majorana energy $E_\textrm{M}$, and by using the corresponding eigenstate of that lowest energy, we get the wave function.
To reduce the finite size effects in determining the Majorana size $\xi_\textrm{M}$ in a zigzag system, we introduce a particle-hole symmetry breaking potential $V \sigma_0 \tau_0$ on one edge, such that one of the Majorana states is pushed away from zero energy.
We then find $\xi_\textrm{M}$ by fitting an exponential to the density of the single Majorana wave function projected on the $x$-axis.
In the straight system we use the eigenvalue decomposition of the translation operator at zero energy~\cite{Nijholt2015} for performance reasons.

\comment{In a straight system, the Majoranas are very poorly localized because of small gap and high velocity}
We show the resulting Majorana wave function densities $\left| \psi_\textrm{M} \right|^2$ in different geometries in Fig.~\ref{fig:wave_functions} using the same Hamiltonian parameter values.
In the straight system [Fig.~\ref{fig:wave_functions}(a)], we see that the decay of the density is long compared to the system size.
The small topological gap combined with the high velocity result in a large Majorana size
\begin{equation}
\label{eq:xi_M}
\xi_\textrm{M}=\hbar \frac{v_\textrm{F}}{E_\textrm{gap}},
\end{equation}
and therefore a minimal topological protection against perturbations.
The wave function extends to the center of the system, resulting in highly overlapping Majoranas and a Majorana coupling $E_\textrm{M}$ comparable to $E_\textrm{gap}$.

\comment{In a zigzag geometry Majoranas are localized within one segment of zigzag independent of details}
We observe that in zigzag systems the Majorana properties improve independent of specific geometric details.
All of the zigzag-type geometries have $\xi_\textrm{M}$ improved by a factor $\sim 70$ and have the Majorana wave function localized within one segment of the zigzag.
Further, the topological gap $E_\textrm{gap}$ is an order of magnitude higher than in the straight junction, and as mentioned in section \ref{sec:band_structures}, the velocity $v_\textrm{F}$ is more than an order of magnitude lower.

%%██████████████████████████████████████████████████████████████████████████
%%██ Topological phase diagram
%%██████████████████████████████████████████████████████████████████████████
\section{Topological phase diagram}

\begin{figure}[!htb]
\includegraphics[width=\columnwidth]{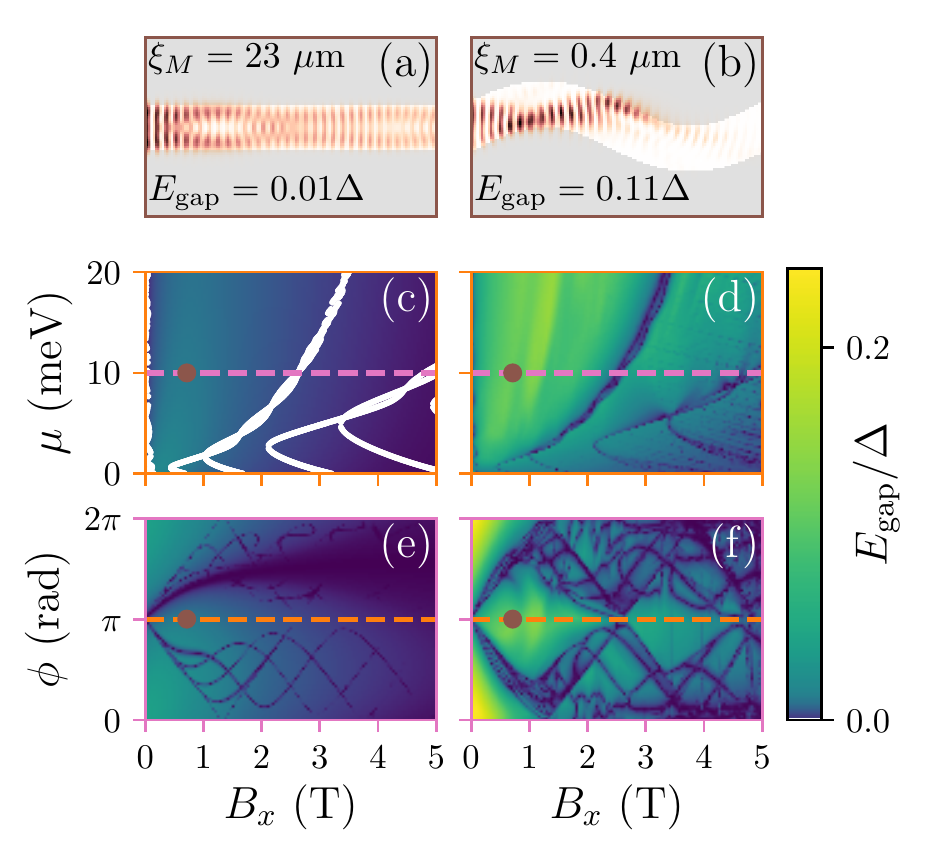}
\caption{A comparison of a straight device (left panels) and a zigzag one (right panels).
The top panels show the Majorana wave functions, near the left edge of the system, at the value of $B_x$ for which $E_\textrm{gap}$ is maximized in a straight geometry for $\mu=\SI{10}{\meV}$ and $\phi=\pi$ as well as the values of the gap and the Majorana size.
The other panels show gap as a function of $\mu$ and $B_x$ at $\phi=\pi$ (middle panels) and as a function of $\phi$ and $B_x$ at $\mu=\SI{10}{\meV}$ (bottom panels).
The dashed lines and the dot indicate the parameters used in the other panels.
Additionally, in subplot (c) we overlay the phase boundaries.
The remaining parameter values are written at the end of Sec.~\ref{sec:setup}, except with $a=\SI{5}{\nm}$ and $L_\textrm{SC}=\SI{800}{\nm}$.
\label{fig:phasediagrams}}
\end{figure}

\comment{We calculate the topological phase diagram using the gap size.}
In Fig.~\ref{fig:phasediagrams} we compare the phase diagrams of the straight and the zigzag junction.
We plot $E_\textrm{gap}$ as a function of magnetic field, chemical potential, $E_\textrm{gap}(B_x, \mu)$; and the superconducting phase difference $E_\textrm{gap}(B_x, \phi)$ for both a straight system [(c) and (e)] and a zigzag system [(d) and (f)].
Additionally, we plot the first $\SI{1300}{\nm}$ (one zigzag period) of the wave functions [(a) and (b)] at the optimal point in parameter space for the straight system.
For the straight system, we calculate $E_\textrm{gap}$ by performing a binary search in $E$ for the energy at which the propagating modes start to appear~\cite{Nijholt2015}.
Additionally, in Fig.~\ref{fig:phasediagrams}(c) we plot the phase boundaries obtained by solving a generalized eigenvalue problem~\cite{Nijholt2015}.
Due to the large size of the zigzag supercell, we are unable to apply these methods to zigzag geometries.
Instead, we calculate $E_\textrm{gap}$ by finding the absolute minimum of the spectrum $E_\textrm{gap}=\min{\left|E(k)\right|}$.
By both observing the gap closings and comparing to the topological phase diagram of the straight system, we then infer the topology of the zigzag system and verify this by calculating the Majorana wave function of a finite length zigzag.

\comment{The introduction of a the zigzag breaks a chiral symmetry, bringing the system from the BDI to the D symmetry class}
The straight system is in the symmetry class BDI \cite{Pientka2017}.
Using the software package Qsymm~\cite{Varjas2018}, we find that the zigzag shape modulation violates the chiral symmetry~\cite{Setiawan2019}, leaving only the particle-hole symmetry and the reflection symmetry with respect to the $x$-axis.

\comment{The phase diagram does not change much, except we see a cleaner spectrum as a result of the D class symmetry.}
Similar to the findings of Pientka \textit{et al.}~\cite{Pientka2017}, we see that the straight geometry has a diamond-shaped topological region.
We also observe additional gap closings due to the BDI symmetry.
The topological phase diagram of the zigzag system has a qualitatively similar shape but a significantly increased topological gap.
The asymmetry of the phase diagram upon replacing $\phi \rightarrow -\phi$ is consistent with the symmetry of the Hamiltonian, because both inversion and time-reversal change both $\phi \rightarrow -\phi$ and $B_x \rightarrow -B_x$.

%%██████████████████████████████████████████████████████████████████████████
%%██ Discussion and Conclusions
%%██████████████████████████████████████████████████████████████████████████
\section{Discussion and Conclusions}

\comment{The zigzag geometry is a controllable tool in hardening the gap and decreasing Majorana size.}
The zigzag geometry increases the topological gap in the high density regime by more than an order of magnitude, as well as substantially reducing of Majorana size.
The improvements occur in a broad range of parameter values, moreover, even using $B_x$ optimal for the straight system in the high density regime (Fig.~\ref{fig:phasediagrams}), the Majorana size $\xi_\textrm{M}$ and $E_\textrm{gap}$ are still more than an order of magnitude better for the zigzag system.
We expect that the improvement of the device performance will significantly simplify the creation of Majorana devices and the detection of Majorana states.
The zigzag geometry offers a controllable way to remove long trajectories, making it easier to rely on than disorder~\cite{Haim2018}, that otherwise has a similar effect.

\comment{Zigzag doesn't do anything against other mechanisms, and therefore it doesn't help to use it ignorantly}
Soft gap may arise due to other mechanisms that do not involve ballistic trajectories: both interface disorder and pair breaking~\cite{Takei2013} or temperature and dissipation~\cite{Liu2017} may create a soft gap.
Further in a multimode junction the mode dependence of transmission~\cite{Stanescu2014} may produce subgap conductance similar to that in a device with a soft gap.
The zigzag geometry has no impact on these alternative phenomena, and it may therefore serve as a tool in distinguishing different mechanisms.

\comment{Current fabrication techniques are compatible with the proposed geometry, and experimental verification can be near.}
Current fabrication techniques are compatible with the proposed geometry; zigzag devices have already been fabricated~\cite{Vries}.
We have demonstrated that the unavoidable variation in the experimental device geometry should not have a detrimental impact on the zigzag devices.

\comment{Optimizing the geometry, more detailed modeling, and a simple analytical estimation are open questions for further research}
Our work is the first demonstration of the impact of the Majorana device geometry on its performance, and it opens a much harder question of finding the optimal geometry.
A promising approach to tackle this question would rely on constructing a quasiclassical model of the zigzag devices.
Finally, we have excluded several important physical effects, such as: disorder, electrostatics, the orbital effect of magnetic field, and the finite thickness of the sample.
While we expect these phenomena not to influence our qualitative findings, a more detailed simulation should provide better guidance to future experiments.

%%██████████████████████████████████████████████████████████████████████████
%%██ Acknowledgments
%%██████████████████████████████████████████████████████████████████████████
\section{Acknowledgments}
We are grateful to  S. Goswami, A. Keselman, P. P. Piskunow, T. Ö. Rosdahl, D. Varjas, F. K. de Vries, Q. Wang, and J. B. Weston for useful discussions.
This work was supported by the Netherlands Organization for Scientific Research (NWO/OCW), as part of the Frontiers of Nanoscience program, two NWO VIDI grants (016.Vidi.189.180 and 680-47-537), and an ERC Starting Grant STATOPINS 638760.

\section{Author contributions}
T.L authored the idea of the zigzag geometry.
B.N. wrote the code and it was extended by T.L.
B.N. performed the writing of the manuscript with input of the other authors.
All authors performed the analysis of the results and planning of the project.

\bibliographystyle{apsrev4-1}
\bibliography{snakemajoranas}
\end{document}